\documentclass[reprint,amsmath,amssymb,aps,pre]{revtex4-2}
\usepackage[utf8]{inputenc}
\usepackage{graphicx}
\usepackage{color}
\usepackage{dcolumn}
\usepackage{bm}
\usepackage{hyperref}
\begin{document}

%\title{A network approach to repurposing drugs. The case of {\color{blue} MYH9 related} nephritis}
\title{Network-based drug repurposing for MYH9-related nephritis}
\author{Muhammad Ali}
\affiliation{DSMN, Ca'Foscari University of Venice, Italy}
\author{Tommaso Gili}
\affiliation{Networks Unit, Scuola IMT Alti Studi Lucca, Italy}
\author{Guido Caldarelli}
\affiliation{Institute of Complex Systems, CNR-ISC, Rome Italy}
\affiliation{DSMN, Ca'Foscari University of Venice, Italy}
\affiliation{London Institute for Mathematical Sciences, Royal Institution, London UK}
\date{\today}
\begin{abstract}
Using tools from network theory, we analyze the organization of a MYH9-oriented drug-like library in chemical space using a multi-descriptor framework. The dataset is drawn from ZINC, a publicly available database of commercially accessible compounds curated for virtual screening and drug discovery. Starting from $6004$ molecules, preprocessing yields $5000$ structurally valid and descriptor-complete compounds.
Similarity is defined via Tanimoto distance on Morgan fingerprints and single-descriptor distances for xLogP, HBD, HBA, molecular weight, and rotatable bonds. For each representation, we construct $k$-nearest-neighbor networks and identify communities using the Louvain-Leiden algorithm. All networks exhibit highly significant modularity ($Q=0.91-0.99$) relative to degree-preserving null models, demonstrating pronounced nonrandom chemical organization across descriptors. 
Cross-descriptor robustness is quantified through a co-clustering matrix over $1.25\times10^7$ molecular pairs, measuring how consistently compound pairs co-occur within the same community across descriptor-specific networks. Although most pairs show limited agreement, a sparse high-consensus core emerges, highlighting the complementarity of the descriptors. 
Minimum spanning trees derived from structural and consensus similarities reveal distinct backbone topologies: a scaffold-driven, sparse structure versus a compact, hub-rich consensus network. Betweenness centrality on these backbones identifies compounds that are both structurally central and descriptor-balanced.
These results provide a statistically validated network representation of chemical space and a principled strategy to extract consensus-stable compounds for downstream screening.

\end{abstract}

\maketitle

\section{Introduction}
Repurposing of drugs identifies new therapeutic uses for approved, failed, or investigational drugs \cite{ashburn2004drug, lamb2006connectivity}. From a complex systems perspective, it poses a challenging question: how to identify functional equivalence between distinct perturbations acting on a highly interconnected and heterogeneous network \cite{barabasi2011network}. Compounds initially developed for different therapeutic indications may elicit similar physiological responses if they interact with the human organism through analogous structural and chemical mechanisms. Quantifying such similarity in a principled and scalable manner remains an open challenge.

Most existing approaches to drug repurposing are based on shared molecular targets, transcriptional signatures, or phenotypic correlations \cite{zhou2020network}. Although effective in specific domains, these strategies often neglect the physical constraints imposed by molecular geometry and chemical composition. However, from a system-level perspective, drug--body interactions are shaped by molecular shape, topology, and physicochemical properties, which determine binding affinities, off-target effects, and the propagation of perturbations across biological interaction networks \cite{baldi2010computational}.

Network theory\cite{caldarelli2007scale,barabasi2016network} provides a natural framework for addressing this problem. Biological organization spans multiple interconnected networks, including protein--protein interaction networks \cite{szklarczyk2015string}, signaling pathways, and organ-level functional dependencies. In this context, drugs can be interpreted as external perturbations whose effects are mediated by their embedding within these networks. Network-based approaches have demonstrated that topological proximity and shared neighborhoods can predict functional similarity and drug efficacy \cite{barabasi2011network,guney2016network,hopkins2008network}.

To integrate structural and chemical information into this framework, it is necessary to move beyond single-layer representations. Multilayer network formalisms allow heterogeneous interaction domains to be encoded simultaneously within a unified mathematical structure \cite{de2013mathematical,battiston2018multiplex}. In such representations, molecular shape similarity, physicochemical distance, and biological interaction profiles can be treated as coupled layers, enabling analysis of how local molecular features influence global network organization.

In this work, we propose a network-theoretic approach to drug repurposing based on the similarity of drug--body interaction topologies. Drugs are represented as nodes in a multilayer network, with weighted edges that encode affinities derived from physical shape and chemical properties. Repurposing opportunities emerge as topological proximity, clustering, or community overlap between drugs associated with distinct therapeutic classes.

By reframing drug repurposing as a problem of similarity and distance in a complex network, this approach shifts the focus from isolated targets to system-level interaction patterns. The resulting framework is naturally aligned with the statistical physics of complex systems and provides a transparent, interpretable basis for identifying candidate drugs, complementary to existing data-driven methodologies \cite{sporns2013structure}.
The specific case study we present here is the exploration of a subset of an extensive database of approved chemical substances (with unknown efficacy), namely the ZINC database (https://zinc.docking.org/), reduced to all compounds that may be used for the treatment of MYH9 nephritis\cite{tabibzadeh2019myh9}. 

MYH9 encodes the heavy chain IIA of non-muscle myosin, a structural component essential for cytoskeletal organization and podocyte stability in renal tissues. Mutations in MYH9 are directly associated with MYH9-related nephropathies, characterized by proteinuria, progressive renal decline, and an increased risk of end-stage kidney disease. Currently, there are no FDA-approved targeted therapies for MYH9-driven nephritis, and clinical management remains largely supportive. Drug-repurposing strategies that exploit existing drug-like chemical libraries offer a viable route to identify molecules capable of stabilizing MYH9-dependent pathways or modulating downstream dysfunction. Therefore, computational prioritization of structurally and physicochemically reliable compounds becomes critical to narrow the search space before biochemical evaluation.

Our approach allows us to explore a curated library of drug-like compounds assembled for nephritis-related therapeutic exploration, with a particular focus on MYH9-associated nephritic disorders. The pipeline combines molecular fingerprints, descriptor-based similarity, community detection, null-model testing, and minimum spanning tree backbones within a unified network-science framework. This activity leverages the principles of complex systems to characterize connectivity patterns, quantify robustness, and identify structurally influential molecules that may serve as candidate compounds for downstream target-specific applications. By integrating complementary molecular representations and grounding them in rigorous statistical methods, we reveal the underlying organization of this MYH9-relevant chemical space and provide a reproducible, interpretable pathway for compound prioritization.

These compounds should ideally reflect robust similarity relationships independent of descriptor choice and therefore represent ideal candidates for subsequent docking, virtual screening, optimization, and repurposing workflows targeting MYH9-associated nephritis.
\section{Results}
With a curated set of $5000$ structurally valid and complete descriptor molecules, the first analytical step was to examine how each descriptor organises chemical space. Because such descriptors capture different chemical principles -- structural motifs, hydrophobicity, polarity, functional groups, size, and conformational freedom -- the networks constructed from them are expected to differ substantially. Such differences determine the types of communities that emerge, the degree to which the descriptors agree, and the need for a consensus analysis. Therefore, each descriptor-specific network offers a distinct view of how the MYH9-focused compound library is organized.

\subsection{Topology}
\begin{figure}
  \centering
  \includegraphics[width=\linewidth]{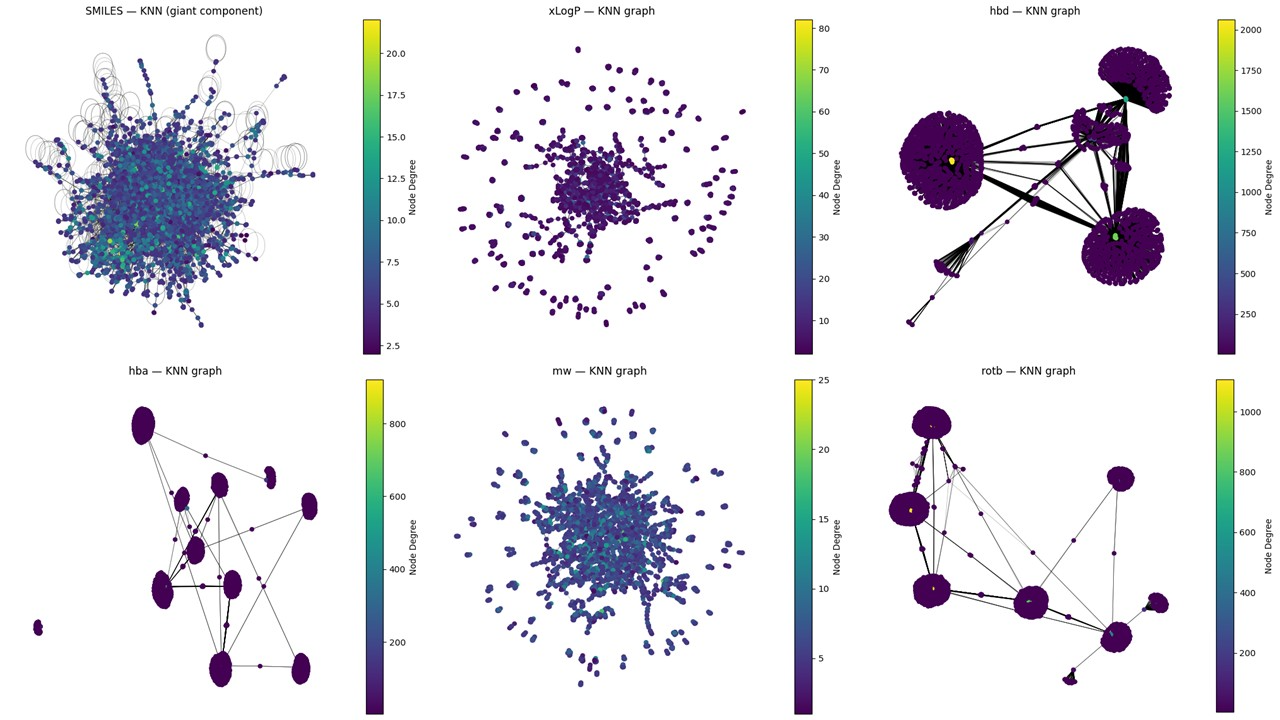}
  \caption{Descriptor-specific KNN networks for SMILES, xLogP, HBD, HBA, MW, and ROTB.}
  \label{fig:knn_networks}
\end{figure}

To observe these descriptor-specific geometries, we constructed six $k$-nearest-neighbor (kNN) similarity networks (Fig.~\ref{fig:knn_networks}), one for each descriptor: SMILES fingerprints, xLogP, HBD, HBA, MW, and ROTB. Each network retains at least $k = 5$ strongest neighbors per node.
Despite identical node sets, the graphs differ in edge density and mesoscale organization, reflecting principles of descriptor-specific organization. Each network contains the same $5000$ nodes, but their connectivity patterns differ because each descriptor encodes chemical similarity differently.
For the SMILES network, the Tanimoto similarity $S_{ij}^{\mathrm{Tan}}$ from Eq.~(\ref{eq:tanimoto}) was used in Morgan fingerprints (radius~2, 2 \,048 bits) to quantify the substructural relatedness among molecules. A force-directed layout shows a dense core with peripheral lobes, reflecting structurally related molecular families.

For descriptor-based networks, we use $z$-score normalization followed by the similarity function of Eq.~(\ref{eq:desc_sim}). The edge counts were: xLogP = 20 \,466; HBD = 26\,961; HBA = 22\,896; MW = 15\,865; ROTB = 24\,760 (Table~\ref{tab:edges_null}). These differences anticipate the observed variation in the community structure and consensus stability between descriptors. The SMILES network had $N=5\,000$ nodes and $E=17\,112$ edges, with the largest connected component containing 4\,962 nodes, average degree $6.844$, average path length $7.21$, and density $0.00137$. Descriptor-based networks showed comparable sparsity but different mesoscale structures.

The SMILES network forms a dense, scaffold-driven structure. The xLogP network forms a compact hydrophobicity-driven core. HBD yields discrete donor-count clusters; HBA shows a functional-group-driven substructure; MW forms continuous gradient windows; ROTB reveals rigid vs flexible modules.

Because each descriptor produces a distinct topology, descriptor-specific community detection becomes essential.

\subsection{Communities}
\begin{figure}
  \centering
  \includegraphics[width=\linewidth]{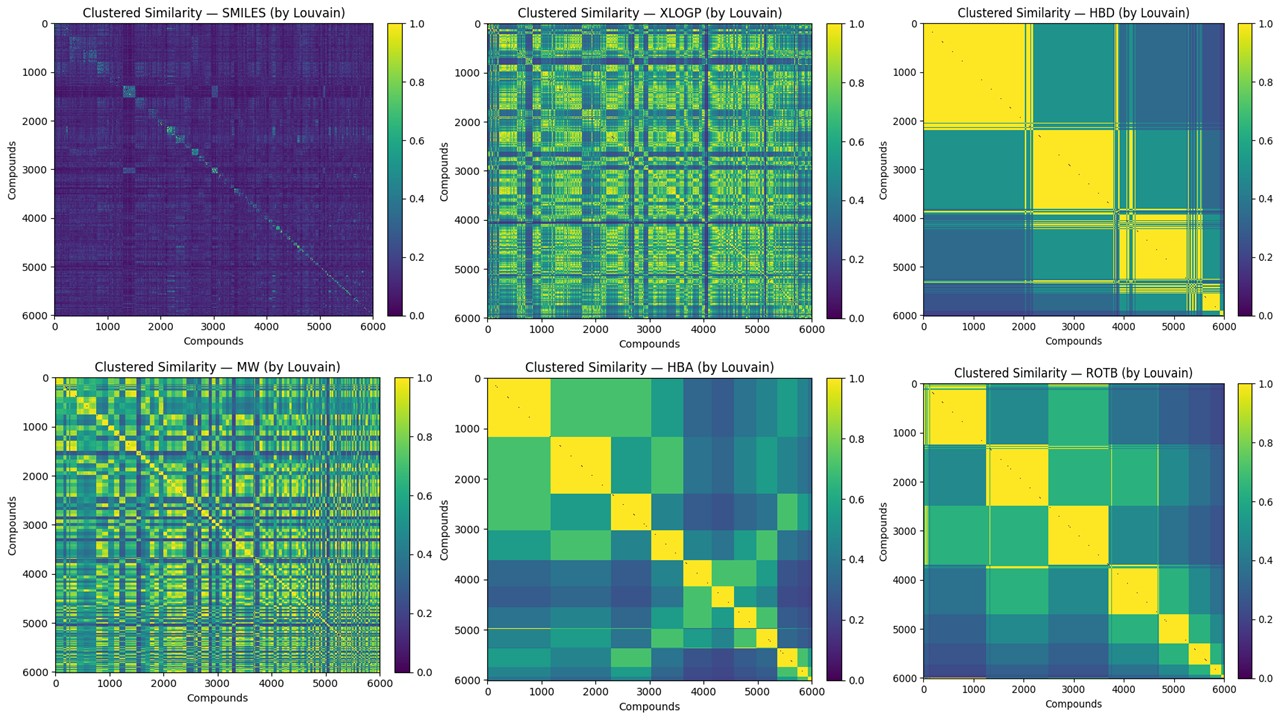}
  \caption{Clustered similarity matrices (SMILES, xLogP, HBD, HBA, MW, ROTB).}
  \label{fig:similarity_matrices}
\end{figure}

We then proceeded by applying the Louvain--Leiden community detection method\cite{blondel2008fast} to each descriptor-specific network. When the vertices are reordered according to their community membership, the adjacency matrices reorganize into a block-diagonal form. Dense diagonal blocks correspond to communities with strong internal connectivity, while off-diagonal regions are sparse. This rearrangement does not alter edge weights or network structure, but only changes the ordering of vertices. The resulting pattern visually confirms the presence of well-defined communities in the network.
These are shown in Figure~\ref{fig:similarity_matrices}. 
SMILES communities reflect scaffold-level diversity, grouping molecules around recurring aromatic frameworks, heterocycles, or fragment motifs (Fig.~\ref{fig:similarity_matrices}), xLogP communities form broader hydrophobicity-based clusters reflecting lipophilicity regimes. HBD and HBA induce functional-group-driven clusters based on hydrogen-bonding capacity. MW produces numerous micro-clusters across narrow mass ranges. ROTB separates rigid molecules from highly flexible ones. This differentiation confirms that chemical similarity depends strongly on the selected descriptor, no single representation adequately captures the chemical space, and descriptor-specific networks uncover complementary molecular relationships. These findings justify the use of multi-view frameworks rather than relying solely on structural or physicochemical models.

% Figure 2

The numbers of communities for the different cases is reported in Table~\ref{tab:communities}.

\begin{table}
\caption{Communities per descriptor.}
\label{tab:communities}
\begin{ruledtabular}
\begin{tabular}{lc}
Descriptor & Communities \\
\hline
SMILES & 98 \\
xLogP  & 149 \\
HBD    & 5 \\
HBA    & 11 \\
MW     & 154 \\
ROTB   & 9 \\
\end{tabular}
\end{ruledtabular}
\end{table}
To determine whether the descriptor-specific communities identified by Louvain clustering reflect genuine chemical structure rather than random graph artifacts, we compared the observed modularity values with those from 50 degree-preserving randomized networks for each descriptor. Original and randomized edge counts are shown in Table~\ref{tab:edges_null}.

\begin{table}
\caption{Original and randomized edge counts for degree-preserving null models.}
\label{tab:edges_null}
\begin{ruledtabular}
\begin{tabular}{lcc}
Descriptor & Original edges & Randomized edges \\
\hline
SMILES & 17\,112 & 17\,110 \\
xLogP  & 20\,466 & 20\,466 \\
HBD    & 26\,961 & 26\,961 \\
HBA    & 22\,896 & 22\,896 \\
MW     & 15\,865 & 15\,865 \\
ROTB   & 24\,760 & 24\,760 \\
\end{tabular}
\end{ruledtabular}
\end{table}

\begin{figure}
  \centering
  \includegraphics[width=\linewidth]{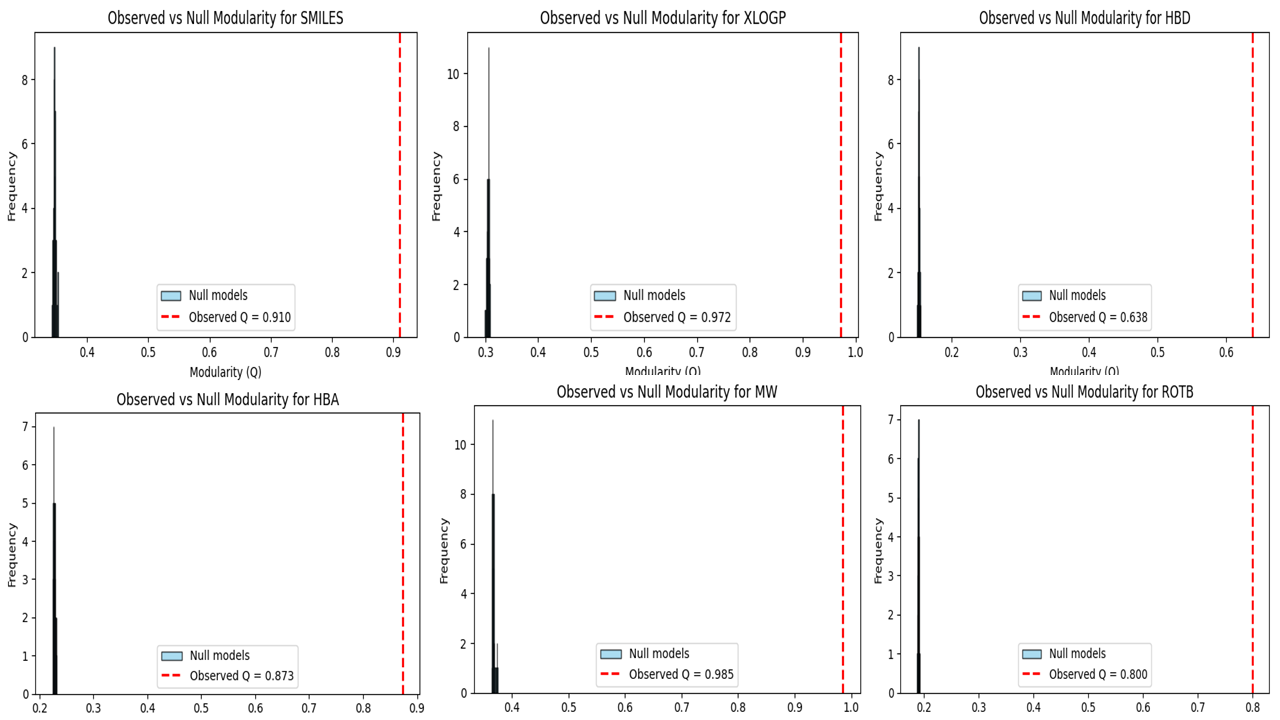}
  \caption{Observed versus null modularity distributions for the six descriptor-specific kNN networks (SMILES, xLogP, HBD, HBA, MW, and ROTB).}
  \label{fig:modularity_dists}
\end{figure}

The distribution of modularity values for the null models was consistently lower than that of the empirical networks. The observed modularity values ($Q_{\text{obs}}$
) were significantly higher (p < 0.001), confirming that the community structures detected were statistically robust and not artifacts of degree correlation. Modularity is defined as:

Figure~\ref{fig:modularity_dists} shows the observed versus null modularity distributions.

% Figure 3

 Figure~\ref{fig:modularity_dists} shows the observed vs null modularity distribution for the six descriptor-specific kNN networks (SMILES, xLogP, HBD, HBA, MW, and ROTB). For each network, the blue histogram shows the modularity distribution of 50 degree-preserving randomized null models, while the red dashed line marks the observed modularity $Q_{\text{obs}}$. In all cases, $Q_{\text{obs}}$ lies far outside the null distribution, indicating a substantial and statistically significant modular organization driven by a descriptor-specific similarity structure.

\begin{table}
\caption{Observed modularity values ($Q$) compared with degree-preserving null models.}
\label{tab:modularity}
\begin{ruledtabular}
\begin{tabular}{lcc}
Network & $Q_{\mathrm{obs}}$ & Null mean $\pm$ SD \\
\hline
SMILES & 0.910 & $0.350 \pm 0.001$ \\
xLogP  & 0.972 & $0.306 \pm 0.002$ \\
HBD    & 0.638 & $0.152 \pm 0.000$ \\
HBA    & 0.873 & $0.228 \pm 0.001$ \\
MW     & 0.985 & $0.366 \pm 0.002$ \\
ROTB   & 0.800 & $0.190 \pm 0.001$ \\
\end{tabular}
\end{ruledtabular}
\end{table}
For each descriptor, none of the 50 degree-preserving null networks achieved a modularity value equal to or greater than the observed value, producing an empirical $p$-value of 0. This confirms that the detected communities are statistically significant and reflect a meaningful chemical organization rather than random partitioning. Because descriptor-level community structures are robust, it is appropriate to examine agreement across descriptors through co-clustering analysis.

\subsection{Cross-descriptor consensus via co-clustering}

To quantify how consistently the six chemical descriptors agree on molecular similarity, we constructed the co-clustering matrix $C$ based on their corresponding community assignments using Eq.~(\ref{eq:co_cluster}). For each descriptor $v$, molecules were clustered independently, and for each molecular pair $(i,j)$, $C_{ij}$ counts the number of descriptors for which molecules $i$ and $j$ appear in the same community. With 5\,000 molecules, the resulting $5\,000 \times 5\,000$ matrix captures all 12\,497\,500 unique molecular pairs, allowing a systematic assessment of the agreement, complementarity, and consensus structure of the descriptor. This analysis was performed over community structures derived from the six kNN-based similarity networks built on SMILES fingerprints, xLogP, HBD, HBA, MW, and ROTB.

\subsubsection{Global distribution of co-clustering values}

\begin{figure}
  \centering
  \includegraphics[width=\linewidth]{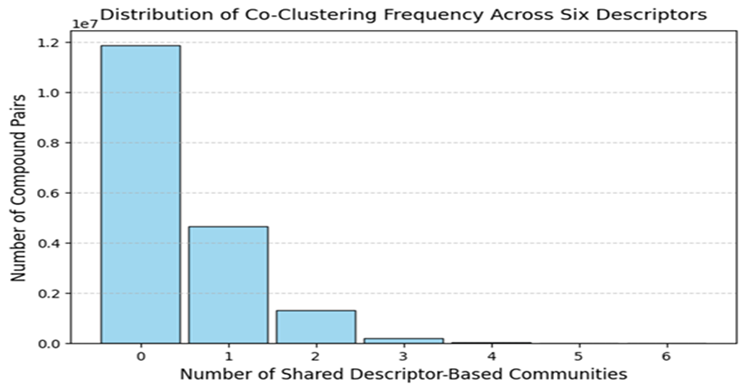}
  \caption{Distribution of co-clustering frequency across 6 descriptors.}
  \label{fig:coclustering_dist}
\end{figure}

The distribution of $C_{ij}$ across all 12.5 million molecular pairs is strongly skewed toward low cross-descriptor agreement. Figure~\ref{fig:coclustering_dist} illustrates the full distribution of co-clustering frequencies; Table~\ref{tab:co_levels} summarizes the exact frequencies.

% Figure 4

Figure 4 illustrates the full distribution of co-clustering frequencies across the 12.5 million molecular pairs. The steep decay shows that most pairs share zero or one descriptor-based community, confirming strong descriptor complementarity. At the same time, the histogram has a long right tail that represent a very small number of high-consensus molecular relationships (shared in 5–6 descriptors), which form robust, descriptor-invariant connections.

% Figure 5
\begin{figure}
  \centering
  \includegraphics[width=\linewidth]{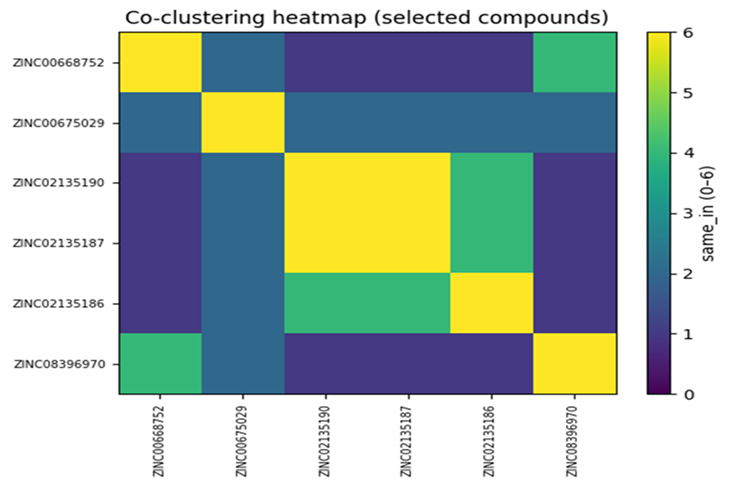}
  \caption{Co-clustering heatmap for a representative subset of molecules.}
  \label{fig:coclustering_heatmap}
\end{figure}

The heatmap in (Fig.~\ref{fig:coclustering_heatmap}) provides a visual understanding of how often selected molecules co-cluster across descriptors. Cells closer to yellow indicate stronger agreement with the descriptor.

\begin{table}
\caption{Number of molecular pairs at each co-clustering level across the six descriptors.}
\label{tab:co_levels}
\begin{ruledtabular}
\begin{tabular}{ccc}
Co-clustering level & Descriptor agreement & Number of pairs \\
\hline
0 & 0 descriptor  & 9\,332\,928 \\
1 & 1 descriptor  & 6\,608\,897 \\
2 & 2 descriptors & 1\,803\,216 \\
3 & 3 descriptors &   248\,689 \\
4 & 4 descriptors &    21\,487 \\
5 & 5 descriptors &     1\,466 \\
6 & All 6 descriptors & 4\,323 \\
\end{tabular}
\end{ruledtabular}
\end{table}

Most molecular pairs fall into the low-agreement range (0--2 descriptors), reflecting significant descriptor diversity. To facilitate interpretation, the levels were grouped into three agreement categories (Table~\ref{tab:co_categories}).

\begin{table}
\caption{Aggregated co-clustering categories.}
\label{tab:co_categories}
\small
\begin{ruledtabular}
\begin{tabular}{llll}
\textbf{Category} &
\textbf{\shortstack[l]{Descriptor\\agreement}} &
\textbf{\shortstack[l]{Number of\\pairs}} &
\textbf{\shortstack[l]{\% of all\\pairs}} \\
\hline
Low agreement &
0--2 descriptors &
17\,745\,041 &
$\approx 99.0\%$ \\
Moderate agreement &
3--4 descriptors &
270\,176 &
$\approx 2.16\%$ \\
High agreement &
5--6 descriptors &
5\,789 &
$\approx 0.046\%$ \\
\end{tabular}
\end{ruledtabular}
\end{table}

Low agreement dominates the dataset, indicating extremely strong complementarity among the descriptors.

\subsubsection{Descriptor Complementarity Overview}

The predominance of low co-clustering frequencies reflects strong complementarity among the six descriptors. Table 6 shows the percentage of pairs at each exact agreement level.
Table~\ref{tab:co_level_props} reports the proportion of molecular pairs at each exact co-clustering level.

\begin{table}
\caption{Proportion of molecular pairs at each exact co-clustering level.}
\label{tab:co_level_props}
\begin{ruledtabular}
\begin{tabular}{cc}
Level & \% of all pairs \\
\hline
0 & 74.7\% \\
1 & 52.9\% \\
2 & 14.4\% \\
3 &  1.99\% \\
4 &  0.17\% \\
5 &  0.011\% \\
6 &  0.034\% \\
\end{tabular}
\end{ruledtabular}
\end{table}

The extremely high proportion of level 0 pairs shows the rarity of strong descriptor overlap. These results confirm that different descriptors emphasize distinct structural or physicochemical dimensions, leading to divergent community assignments.

\subsubsection{Per-compound descriptor stability}

To examine molecular behaviour at the compound level, we computed three metrics for each of the 5\,000 molecules.

\paragraph{Unique Cluster Membership:}

First, the number of distinct cluster labels a molecule $i$ receives across descriptors is
\begin{equation}
U(i) = \left| \big\{ c_i^{(v)} : v = 1,\dots,V \big\} \right|.
\label{eq:unique_clusters}
\end{equation}

\paragraph{Number of Molecules Sharing $\geq$ 1
  Cluster:}
Second, the number of molecules sharing at least one cluster with $i$ is
\begin{equation}
N_{\ge 1}(i) = \big| \{ j : C_{ij} \ge 1 \} \big|.
\label{eq:n_ge1}
\end{equation}
\paragraph{Total Co-Clustering Strength:}
Third, the global descriptor agreement for molecule $i$ is given by the total co-clustering strength
\begin{equation}
S_{\mathrm{tot}}(i) = \sum_j C_{ij}.
\label{eq:total_c}
\end{equation}

Illustrative examples for five molecules are reported in Tables~\ref{tab:unique_clusters}--\ref{tab:co_strength}.

\begin{table}
\caption{Unique cluster membership across descriptors (example 5 compounds).}
\label{tab:unique_clusters}
\begin{ruledtabular}
\begin{tabular}{lc}
ZINC ID      & Unique cluster count \\
\hline
ZINC00668752 & 4 \\
ZINC00675029 & 3 \\
ZINC02135190 & 5 \\
ZINC02135187 & 5 \\
ZINC02135186 & 4 \\
\end{tabular}
\end{ruledtabular}
\end{table}

Unique cluster counts indicate how consistently each molecule is placed across descriptors. Most molecules fall between 3 and 5 clusters, indicating moderate descriptor coherence:
3 clusters reflect strong agreement and stable neighbourhoods, while 5–6 clusters indicate increasing disagreement among descriptors and multimodal behaviour.

\begin{table}
\caption{Number of molecules sharing at least one community with each example compound.}
\label{tab:share_cluster}
\begin{ruledtabular}
\begin{tabular}{lc}
ZINC ID      & Molecules sharing $\geq 1$ cluster \\
\hline
ZINC00668752 & 2\,917 \\
ZINC00675029 & 3\,073 \\
ZINC02135190 & 3\,013 \\
ZINC02135187 & 3\,013 \\
ZINC02135186 & 2\,968 \\
\end{tabular}
\end{ruledtabular}
\end{table}

Molecules typically share a cluster with ~3,000 others, reflecting a densely connected chemical. Each molecule shares a cluster with ~2,900–3,100 others, demonstrating that the chemical space is highly interconnected even before enforcing consensus. space.

\begin{table}
\caption{Total co-clustering strength for example molecules.}

\label{tab:co_strength}
\begin{ruledtabular}
\begin{tabular}{lc}
ZINC ID      & Total co-clustering strength \\
\hline
ZINC00668752 & 4\,133 \\
ZINC00675029 & 4\,093 \\
ZINC02135190 & 4\,028 \\
ZINC02135187 & 4\,028 \\
ZINC02135186 & 3\,969 \\

\end{tabular}
\end{ruledtabular}
\end{table}

Total co-clustering strength measures global consensus for each molecule across all descriptors. Higher values correspond to molecules with strong, consistent similarity patterns across the descriptors. Lower values indicate descriptor-specific or inconsistent clustering behaviour.
\subsubsection{Consolidated Interpretation}

Table~\ref{tab:co_summary} consolidates the main descriptor-based co-clustering insights.

\begin{table}
\caption{Summary of descriptor-based co-clustering insights.}
\label{tab:co_summary}
\small
\begin{ruledtabular}
\begin{tabular}{@{}l l@{}}
\textbf{Aspect} &
\textbf{\parbox[t]{0.54\columnwidth}{\raggedright Key observation}} \\
Descriptor complementarity &
\parbox[t]{0.54\columnwidth}{\raggedright $\approx 99\%$ of molecular pairs agree in only 0--2 descriptors.} \\
Moderate consensus &
\parbox[t]{0.54\columnwidth}{\raggedright 2.16\% of pairs align in 3--4 descriptors.} \\
High consensus core &
\parbox[t]{0.54\columnwidth}{\raggedright Only 0.046\% of pairs agree in 5--6 descriptors.} \\
Unique cluster stability &
\parbox[t]{0.54\columnwidth}{\raggedright Most molecules appear in 3--5 clusters across descriptors.} \\
Descriptor interconnectedness &
\parbox[t]{0.54\columnwidth}{\raggedright Each molecule co-clusters with roughly 3,000 others in at least one descriptor.} \\
Global co-clustering strength &
\parbox[t]{0.54\columnwidth}{\raggedright Values typically range from 3,900--4,200 across molecules.} \\
\end{tabular}
\end{ruledtabular}
\end{table}

The combined analysis of descriptor-based co-clustering reveals a chemically diverse, multimodal landscape in which different descriptors capture distinct structural and physicochemical relationships. The overwhelmingly high proportion of low-agreement molecular pairs demonstrates that these descriptors encode largely complementary information. Despite this heterogeneity, a small but meaningful subset of molecules consistently aligns across multiple descriptors, forming a stable consensus core that reflects shared underlying chemical principles.

These results highlight the importance of using multi-descriptor frameworks for chemical space exploration and drug repurposing. By integrating neighborhoods derived from structurally distinct descriptors, one can obtain a richer, more informative representation of molecular similarity. This enhanced view helps identify robust chemical families, reveals stable connectivity patterns, and improves the reliability of downstream tasks such as target prediction, clustering, and screening. Overall, the co-clustering analysis demonstrates both the diversity and the hidden coherence within the chemical landscape, reinforcing the value of multi-perspective molecular characterization.

 The co-clustering results show that only 0.046\% of molecular pairs achieve agreement across five or six descriptors, forming a small but highly stable consensus core. These high-consensus relationships are statistically rare yet chemically robust, reflecting descriptor-independent similarity. Because the consensus MST aims to preserve only the most reliable cross-descriptor relationships, these high-agreement molecular pairs naturally serve as structural anchors. Their consistent behavior across multiple chemical perspectives makes them ideal backbone edges for constructing a unified, consensus-based topology.

\subsection{Structural and consensus backbones via minimum spanning trees}

\begin{figure}
  \centering
  \includegraphics[width=\linewidth]{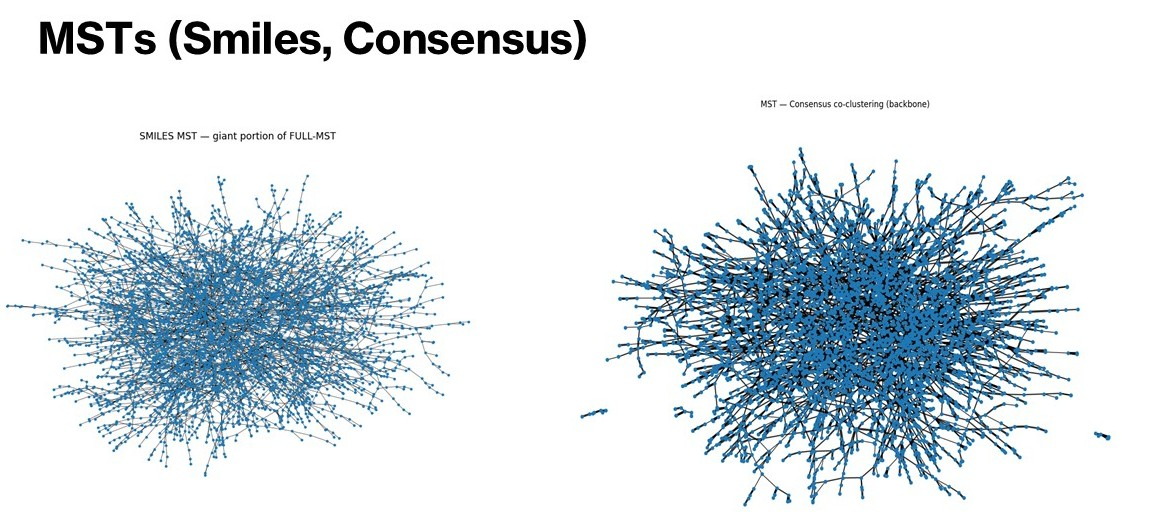}
  \caption{SMILES MST vs consensus MST.}
  \label{fig:msts}
\end{figure}

To derive global, interpretable backbones of chemical space under both structural and consensus similarity, we constructed two MSTs: one based on SMILES-derived structural similarity, and one based on consensus similarity aggregated from all six descriptors (Fig.~\ref{fig:msts}). Distances for both MSTs were computed as in Eq.~(\ref{eq:dist}), where $S_{ij}$ denotes the pairwise similarity. For the consensus MST, similarities were defined by Eq.~(\ref{eq:s_cons}), with $C_{ij}$ representing the number of times molecules $i$ and $j$ co-clustered across the six descriptor-specific networks.

% Figure 6

The SMILES-based MST consisted of 4\,972 nodes and 4\,972 edges, with 2\,847 leaf nodes (57.2\%). The maximum degree was~9, the diameter reached~43, and the average path length was~17.4. This topology reflects a scaffold-dominated organization, characterized by elongated chains of structural variants connected by relatively sparse bridging compounds.

In contrast, the consensus MST contained 4\,995 nodes and 4\,995 edges, including 2\,601 leaf nodes (52.1\%). It exhibited a higher maximum degree (12), a reduced diameter (31), and a shorter average path length (12.8). These properties indicate a more compact and integrated backbone, where molecules consistently grouped across multiple descriptors form a tighter, more cohesive chemical neighbourhood.

\begin{table}
\caption{Global properties of SMILES-based and consensus minimum spanning trees (MSTs).}
\label{tab:mst_properties}
\begin{ruledtabular}
\begin{tabular}{lcc}
Property & SMILES-based MST & Consensus MST \\
\hline
Nodes               & 4\,972 & 4\,995 \\
Edges               & 4\,972 & 4\,995 \\
Leaf nodes          & 2\,847 (57.2\%) & 2\,601 (52.1\%) \\
Maximum degree      & 9     & 12 \\
Diameter            & 43    & 31 \\
Average path length & 17.4  & 12.8 \\
\end{tabular}
\end{ruledtabular}
\end{table}
Taken together, the SMILES MST and consensus MST reveal two distinct connectivity regimes: structural adjacency and multi-descriptor chemical coherence. These form the foundation for identifying central molecular hubs.

\subsection{Identification of backbone key compounds}

To determine which molecules play the most influential bridging roles within the structural and consensus MSTs, we computed betweenness centrality for every node using Eq.~(\ref{eq:bc}). Betweenness centrality quantifies how often a molecule lies along the shortest paths between pairs of other molecules. High-centrality molecules serve as critical connectors between chemical families, and their importance differs depending on whether structural or consensus similarity is used.

Table~\ref{tab:bc_summary} summarises betweenness-centrality statistics for the top 50 nodes in each MST.

\begin{table}
\caption{Summary statistics of betweenness centrality for top 50 nodes in each MST.}
\label{tab:bc_summary}
\begin{ruledtabular}
\begin{tabular}{lcccccc}
Network & Count & Mean & Median & SD & Min & Max \\
\hline
Consensus MST & 50 & 0.3743 & 0.4092 & 0.0821 & 0.2352 & 0.6557 \\
SMILES MST    & 50 & 0.3738 & 0.3516 & 0.0791 & 0.2540 & 0.5279 \\
\end{tabular}
\end{ruledtabular}
\end{table}

The top 10 high-centrality compounds in the SMILES MST are reported in Table~\ref{tab:top10_smiles}.

\begin{table}
\caption{Top 10 high-centrality compounds in the SMILES MST.}
\label{tab:top10_smiles}
\begin{ruledtabular}
\begin{tabular}{ccc}
Rank & Compound ID & Betweenness \\
\hline
1  & ZINC00670787 & 0.527908 \\
2  & ZINC01895029 & 0.517964 \\
3  & ZINC08399011 & 0.503845 \\
4  & ZINC00674567 & 0.486375 \\
5  & ZINC02134860 & 0.469088 \\
6  & ZINC00670791 & 0.463146 \\
7  & ZINC02134862 & 0.459930 \\
8  & ZINC02134855 & 0.458765 \\
9  & ZINC08399014 & 0.458300 \\
10 & ZINC00675029 & 0.454215 \\
\end{tabular}
\end{ruledtabular}
\end{table}

These molecules form key structural connectors in scaffold space, bridging otherwise distant structural motifs.

The top 10 high-centrality compounds in the consensus MST are summarised in Table~\ref{tab:top10_consensus}.

\begin{table}
\caption{Top 10 high-centrality compounds in the consensus MST.}
\label{tab:top10_consensus}
\begin{ruledtabular}
\begin{tabular}{ccc}
Rank & Compound ID & Betweenness \\
\hline
1  & ZINC00662981 & 0.655728 \\
2  & ZINC02101627 & 0.486345 \\
3  & ZINC00995741 & 0.462674 \\
4  & ZINC08399817 & 0.461278 \\
5  & ZINC08399809 & 0.452311 \\
6  & ZINC00655084 & 0.442167 \\
7  & ZINC00675029 & 0.438614 \\
8  & ZINC08399990 & 0.436192 \\
9  & ZINC00668752 & 0.434819 \\
10 & ZINC08396970 & 0.431860 \\
\end{tabular}
\end{ruledtabular}
\end{table}

In contrast to the SMILES MST hubs, these consensus hubs represent molecules that exhibit multi-descriptor coherence, aligning across structural, functional, hydrophobic, and conformational dimensions. Together, the two hub sets provide a dual view of chemical importance within the MYH9-oriented library: one reflecting pure structural transitions in scaffold space, the other reflecting balanced multi-parameter chemical similarity. The latter, in particular, highlights consensus-stable compounds that are well suited as prioritised candidates for subsequent MYH9-focused docking, virtual screening, or repurposing studies.

\section{Discussion}

This work established a comprehensive multi-descriptor analysis framework for characterizing the structural organization of a MYH9-oriented drug-like chemical space, integrating similarity modeling, network representations, community detection, null-model validation, co-clustering, and backbone extraction. By systematically comparing descriptor-specific similarity networks and identifying consensus-stable molecular relationships, we demonstrated how chemically meaningful structure emerges across multiple representational perspectives within a therapeutically relevant library. The resulting findings provide a reproducible foundation for interpreting molecular organization and identifying compounds that play a central role in the similarity landscape of a MYH9-focused repurposing set.

We observed that structural diversity manifests itself differently in the six descriptor-specific networks. The SMILES-based network forms a dense scaffold-driven space characterized by extensive structural transitions among aromatic frameworks, heterocycles, and hybrid scaffolds. In contrast, descriptor-derived networks —particularly HBD and ROTB—produce discrete partitions that reflect categorical chemical counts rather than continuous gradients. xLogP, MW, and HBA exhibit intermediate behavior, in which molecules form cohesive neighborhoods based on hydrophobicity or sizes. These observations confirm that chemical similarity is strongly representation-dependent, that no single descriptor captures the entire landscape, and that descriptor-specific networks uncover complementary molecular relationships.

Community detection applied to each descriptor-specific network reveals clear mesoscale organization, with distinct patterns emerging across descriptors. Structural fingerprints (SMILES), MW, and xLogP produce highly granular partitions that reflect subtle and continuous gradients in molecular variation. Conversely, discrete descriptors yield fewer chemically interpretable communities tied to donor/acceptor counts or conformational flexibility. Degree-preserving null models consistently yield substantially lower modularity distributions than observed values, demonstrating that the identified communities are not artifacts of network topology or density. This statistical validation confirms that the detected clusters correspond to reproducible, chemically meaningful similarity relationships rather than to algorithmic bias.

The co-clustering analysis reveals a striking observation: most molecular pairs agree on only a few descriptors. Approximately $99\%$ pairs match in zero, one, or two descriptors, highlighting strong representational complementarity across molecular features. Only $\sim 0.046\%$ pairs exhibit agreement on five or six descriptors, forming a small but extremely stable consensus core. These high-agreement relationships represent descriptor-independent similarities, indicating molecular pairs that consistently appear close in structural, hydrophobic, bonding, and flexibility scales. This subset serves as the foundation for a consensus-based network backbone.

Minimum spanning trees expose backbone connectivity under both structural (SMILES-based) and consensus similarity. The SMILES-based MST reflects a scaffold-driven structure with long branches, limited hubs, and large diameter, illustrating gradual structural transitions. The consensus MST is more compact, with higher-degree hubs and shorter path lengths because multi-descriptor-consistent molecules act as central connectors. This duality is critical for drug discovery: scaffold hopping emerges naturally from the structural backbone, while multi-parameter optimization is reflected in the consensus backbone. Ideally, drug-like molecules lie near consensus hubs.

Betweenness centrality further highlights the distinct classes of central molecules. Structural-MST central compounds bridge scaffold clusters and link remote regions of chemical space. The Consensus-MST central compounds are multi-property balanced molecules capable of connecting descriptor-stable regions. Such molecules are ideal candidates for prioritizing balanced chemical profiles or representative scaffolds and may serve as lead structures for subsequent screening or analogue development.

Finally, it is important to emphasize that this study focuses on structural organization and consensus-driven prioritization rather than direct prediction of biological interactions. Docking-based interaction assessment and biological validation (e.g., podocyte-specific assays, MYH9-dependent pathway readouts) constitute natural downstream steps once a statistically justified subset of candidates has been identified. By constructing an interpretable and robust chemical-space backbone, the present framework provides a rational pre-filter for subsequent target-specific screening pipelines.
\section{conclusion}

This study presented an integrated chemical-space analysis framework tailored to a collection of MYH9-relevant drug-like compounds. Using multiple molecular representations and network-science principles, we systematically identified descriptor-stable relationships, statistically meaningful community structures, and backbone connectivity patterns that reveal how molecules organize within this therapeutic search space. The comparison between structural and consensus-based networks demonstrated that only a very small subset of molecules consistently align across multiple descriptors, identifying a chemically robust core that forms reliable connectivity backbones. These consensus hubs, together with their high centrality in the network, provide interpretable indicators of balanced chemical similarity and structural representativeness, thus offering rational entry points for downstream drug-repurposing workflows related to MYH9-associated nephritis.

Importantly, the prioritization provided here serves as a necessary intermediate computational stage before targeted-specific screening or biological evaluation. Rather than performing unbiased docking across thousands of compounds, our pipeline reduces chemical redundancy, highlights meaningful scaffold transitions, and isolates compounds that exhibit reproducible multi-parameter coherence. These results establish an informed foundation for target-oriented modeling, including structure-based screening, pharmacophore refinement, network pharmacology, and pathway-level validation of MYH9-modulating candidates. In general, this work demonstrates that structurally grounded, statistically validated backbone extraction can serve as a powerful filtering and prioritization layer in repurposing pipelines targeting rare or underexplored nephropathic mechanisms, including MYH9-associated pathological phenotypes.

\section{Methods}
\label{s:Methodology}
\subsection{Data preprocessing and descriptor resolution}
 To ensure uniform downstream analysis across all similarity spaces, descriptor headers were normalized through a robust synonym-matching routine. When any of the five descriptors were missing, but a valid SMILES was available, we computed the values directly from the structure using RDKit (Crippen logP for xLogP; Lipinski donor/acceptor counts; and exact molecular weight). Molecules with invalid or unparsable SMILES strings and rows lacking essential descriptors were removed from the working set. This quality-control step led to a consistent index of 5,000 chemically valid entries across all network analyses. This reduction from $6,004$ to $5,000$ compounds is expected in cheminformatics workflows when fingerprints or descriptor computations fail for a subset of structures; keeping a standard index enables one-to-one comparison between descriptor-specific networks and prevents missing-node artifacts in integrated analyses.

\subsection{Network Definition}
Starting from the above data
%the adjacency matrices encode the explicit network structure obtained after transforming molecular similarity information into sparse similarity graphs. E
 each {\bf Node} in the graph corresponds to a single compound in the dataset, identified by its ZINC ID, and the node set is identical across all networks considered. 
 {\bf Edges} vary in their definition, but always represent a similarity relationships between nodes defined through a $k$-nearest-neighbor (k-NN) graph construction. We made two different methods for two different questions. In one case, we considered the structural properties; in the second, the physico-chemical properties.

For the second case, the edges encode similarity along a single physicochemical dimension (defined below) rather than structural overlap. Two compounds are connected if they are among each other’s nearest neighbours in a z-score–normalized descriptor space, with edge weights reflecting numerical proximity in that specific property. 
 
\subsection{Networks from molecular structure}
For the first case, starting from the SMILES coding of a compound, we considered its {\em Morgan fingerprints} (see below) and created an edge between nodes $i,j$ when the Tanimoto/Jaccard similarity $S_{ij}$ was defined as

\begin{equation}
S_{ij}^{\mathrm{Tan}}
= \frac{\lvert A_i \cap A_j \rvert} {\lvert A_i \cup A_j \rvert} =\frac{\lvert A_i \cap A_j \rvert}
       {\lvert A_i \rvert + \lvert A_j \rvert - \lvert A_i \cap A_j \rvert},
\label{eq:tanimoto}
\end{equation}
is above a certain threshold; the edge weights equal the corresponding Tanimoto coefficient\cite{jaccard1912distribution,tanimoto1958elementary} on circular Morgan fingerprints (radius 2, 2048 bits) to quantify substructural relatedness between molecules. From the full similarity matrix, a K-nearest-neighbor (KNN) graph with $K = 5$ was constructed to retain the most informative edges while ensuring sparse, interpretable topology. Edge weights encode similarity in $[0, 1]$. The resulting network contains $5,000$ nodes and $17,112$ weighted edges. A force-directed layout reveals a dense, cohesive core and several peripheral lobes, consistent with families of structurally related compounds linked by strong intra-family edges and bridged by weaker inter-family ties. 
 
{\bf Morgan fingerprints} are a widely used molecular representation in cheminformatics designed to encode the topological structure of a molecule into a fixed-length numerical vector, enabling efficient comparison, similarity search, and machine-learning analysis. They belong to the family of circular fingerprints and are most commonly known for their implementation as Extended Connectivity Fingerprints (ECFPs) \cite{rogers2010extended}.
The core idea behind Morgan fingerprints is to represent a molecule by systematically characterizing the local atomic environments around each atom. Initially, each atom is assigned an identifier based on basic atomic properties such as atomic number, valence, charge, and aromaticity. These identifiers are then iteratively updated by aggregating information from neighboring atoms, effectively expanding the atomic environment to increasing radii. At each iteration, the resulting substructures are encoded and mapped, typically via hashing, into a high-dimensional binary or count-based vector. The final fingerprint captures the presence of specific substructural patterns within the molecule, reflecting its connectivity and chemical functionality..
 
%For descriptor-based networks, edges connect compounds that are nearest neighbors in the one-dimensional, $z$-score--normalized descriptor space, with edge weights defined by the similarity transformation given in Eq.~(2).In all cases, sparsity is enforced implicitly by the k-NN rule: only the strongest local similarity relations for each compound are retained, while all other pairwise similarities are set to zero in the adjacency matrix. As a result, the adjacency matrices provide a direct, topological view of how different similarity definitions shape local connectivity patterns in chemical space, independently of any higher-order clustering or consensus analysis.
%An entry $A_{ij}$ in the adjacency matrix is nonzero only if a direct edge exists between compounds $i$ and $j$ in the corresponding similarity network, and its value equals the weight of that edge. By construction, the diagonal entries are zero, as self-loops are excluded.

\subsection{Networks from Descriptor-based networks}
\label{s:descriptors}
To complement structural similarity, we constructed five descriptor-specific KNN networks using z-score normalization followed by a one-dimensional similarity function
\begin{equation}
S_{ij} = \frac 1 {(1 + 
|z_i - z_j|)}
\label{eq:desc_sim}
\end{equation}
where $z_i$ and $z_j$ are z-score normalized descriptor values for compounds $i$ and $j$. This transformation scales molecular similarity between $0$ and $1$, allowing uniform comparison across descriptors.

\begin{itemize}
\item {\bf $x\log P$} that captures lipophilicity; the network shows multiple moderately dense regions with hubs indicative of lipophilic series. Formally it is defined as 
$x\log P = \log_{10}(c_{\mathrm{o}}/c_{\mathrm{w}})$, 
where $c_o$ and $c_w$ denote the equilibrium concentrations in octanol and water, respectively.

\item {\bf HBD} that is Hydrogen Bond Donors. It is a molecular descriptor that counts the number of functional groups in a molecule capable of donating a hydrogen atom in a hydrogen bond. Typical HBD groups include: OH (alcohols, phenols); -NH, -NH$_2$ (amines, amides);
–SH (thiols, less commonly). The quantity is integer-valued. The network exhibits a small number of large, cohesive communities driven by donor count similarity.
\item{\bf HBA} that is Hydrogen Bond Acceptor, It is a molecular descriptor that counts the number of atoms in a molecule capable of accepting a hydrogen bond, typically atoms with lone pairs. Common HBA atoms/groups include: Oxygen atoms (e.g., carbonyl O, ether O, hydroxyl O); Nitrogen atoms (e.g., amines, imines, nitriles); Sulfur atoms (e.g., thioethers, sulfones) that forms moderately fragmented mesoscale structure, consistent with acceptor-count diversity across the library.
\item{\bf Molecular weight} represents the size and mass of a molecule, and it serves as a proxy for how easily the compound can cross biological membranes. We can assume that $MW \le 500 Da$ is an empirical upper bound for compounds that tend to show acceptable oral absorption.
Indeed, large values of $MW$ are connected with larger polar surface area, increased steric bulk, greater flexibility (often more rotatable bonds, see below) that produce lower passive membrane permeability, poorer aqueous solubility, higher conformational entropy penalties upon binding. These factors collectively reduce the likelihood that a molecule can diffuse through lipid bilayers.
\item{\bf Rotatable-bond counts} Introduced as refinement of the above four parameters \cite{veber2002molecular}, this quantity takes into account the conformational flexibility and allows to distinguish between otherwise similar molecules. Indeed, two molecules with identical MW and logP can behave very differently if one is rigid and the other highly flexible.

\end{itemize}
The five descriptors described above are taken from the (generally accepted) Lipinski's rule of five\cite{lipinski2012experimental}, that is "A compound is more likely to be orally drug-like if it does not violate more than one of the following criteria":
(i) Hydrogen bond donors (HBD) $\le$ 5; (ii) Hydrogen bond acceptors (HBA) $\le$  10; (iii) Molecular weight (MW) $\le 500\  Da$; (iv) Octanol–water partition coefficient (xlogP) $\le$ 5

Each network retains $K = 5$ strongest neighbors per node. Despite identical node sets, the graphs differ in edge density and mesoscale organization, reflecting descriptor-specific organization principles. Edge counts were: xLogP $= 20,466$; HBD $= 26,961$; HBA $= 22,896$; MW $= 15,865$; ROTB $= 24,760$. These differences anticipate the observed variation in community structure and consensus stability across descriptors.

\subsection{Modularity and Louvain method of computing }
The modularity (Q) of a network, which measures the density of links within communities compared to random expectation, is defined as:

\begin{equation}Q = \frac{1}{2m} \sum_{i,j}\left[A_{ij} - \frac{k_i k_j}{2m}\right]\delta(c_i, c_j)\end{equation}

where $A_{ij}$ is the edge weight between nodes $i$ and $j$, $k_i$ and $k_j$ are their degrees, and $\delta(c_i, c_j)$ equals $1$ if both belong to the same community. Higher modularity values indicate stronger community structures.

The Louvain algorithm\cite{blondel2008fast} (one of the most widely used methods for community detection in large networks, owing to its computational efficiency and its ability to uncover meaningful mesoscopic structure) is based on recursive modularity optimisation.  The algorithm proceeds iteratively in two main phases: first, each node is assigned to its own community and nodes are then greedily reassigned to neighboring communities if this results in an increase in modularity; second, the identified communities are aggregated into super-nodes, yielding a reduced network on which the process is repeated. This hierarchical aggregation continues until no further modularity improvement is possible. As a result, the Louvain algorithm naturally produces a multilevel, hierarchical partition of the network and scales efficiently to networks with millions of nodes, making it particularly suitable for the analysis of large, complex systems where community structure plays a central role.

\subsection{Null models and modularity validation}

Degree-preserving randomized networks were generated via double-edge swaps.\cite{maslov2002specificity} For each null, modularity was recomputed, yielding a distribution $\{Q_r\}$. Statistical significance was estimated as
\begin{equation}
Z_Q = \frac{Q_{\mathrm{obs}} - \mu_{Q_r}}{\sigma_{Q_r}},
\label{eq:zq}
\end{equation}
where $Q_{\mathrm{obs}}$ is the observed modularity and $\mu_{Q_r}$ and $\sigma_{Q_r}$ are the mean and standard deviation of modularity over the null ensemble.\cite{cimini2019statistical} Degree-preserving null models isolated the contribution of the empirical topology from that of the degree distribution, providing a rigorous baseline for modularity significance and following principles of statistical network comparison widely used in complex network theory.\cite{caldarelli2007scale,cimini2019statistical}

\subsection{Cross-view consensus via co-clustering}

A co-clustering matrix was constructed across multiple network views. For each descriptor $v$ and each molecular pair $(i,j)$, we computed
\begin{equation}
C_{ij} = \sum_{v=1}^{V}
\mathbf{1}\!\big(c_i^{(v)} = c_j^{(v)}\big),
\label{eq:co_cluster}
\end{equation}
where $c_i^{(v)}$ is the community of compound $i$ in view $v$, $V = 6$ is the number of descriptors, and $\mathbf{1}(\cdot)$ is the indicator function. This matrix quantified how consistently molecule pairs co-occurred in the same community across different similarity definitions.\cite{monti2003consensus,lancichinetti2012consensus,caldarelli2012networks}

\subsection{Minimal Spanning Trees}
The Minimum Spanning Tree (MST) captures the backbone of each network by connecting all nodes with minimal total distance. Indeed, given a connected graph with weighted edges, an MST is defined as the acyclic subgraph that connects all nodes while minimizing the total edge weight. By construction, the MST contains exactly $N-1$ edges for a network of N nodes, preserving global connectivity while discarding redundant or weaker links. This property makes MSTs particularly valuable for analyzing complex systems in which the full network is dense, noisy, or difficult to interpret, as the MST highlights the most significant relationships encoded in the weight structure while removing unnecessary links \cite{barabasi2016network,caldarelli2007scale}.
The structure is obtained by ranking the edges according to their value of similarities and connecting nodes from the most similar couple of edges. Whenever an edge creation would form a loop, these move is canceled and we pass to consider the next couple of nodes. The procedure stops when all the nodes are connected.  For backbone analysis, similarities were converted into distances,
\begin{equation}
d_{ij} = 1 - S_{ij},
\label{eq:dist}
\end{equation}
and to get the minimum spanning trees (MSTs) we used Kruskal’s algorithm \cite{kruskal1956shortest,graham1985history}. For the consensus MST, similarities were defined as
\begin{equation}
S_{ij}^{\mathrm{cons}} = \frac{C_{ij}}{V},
\qquad
d_{ij}^{\mathrm{cons}} = 1 - S_{ij}^{\mathrm{cons}},
\label{eq:s_cons}
\end{equation}
with $C_{ij}$ from Eq.~(\ref{eq:co_cluster}) representing the number of times molecules $i$ and $j$ co-clustered across the six descriptor-specific networks.

\subsection{Network diagnostics and candidate prioritization}

Network diagnostics included degree distributions, path lengths, diameters, edge-weight statistics, and leaf ratios, aligning with network-analysis frameworks in chemical-space mapping.\cite{scalfani2022chemical} Cross-view agreement was measured using edge-set Jaccard overlap and degree correlations.

Candidate prioritization utilized betweenness centrality,\cite{freeman1977set}
\begin{equation}
BC(v) =
\sum_{s \neq v \neq t}
\frac{\sigma_{st}(v)}{\sigma_{st}},
\label{eq:bc}
\end{equation}
where $\sigma_{st}(v)$ is the count of shortest paths between nodes $s$ and $t$ that pass through node $v$. Nodes with high betweenness typically occupy bridging positions between modules and therefore act as mediators of structural or functional diversity.\cite{barabasi2009scale,caldarelli2012networks}

\subsection{Consensus network integration and robustness analysis}

Similarity Network Fusion (SNF)\cite{wang2014similarity} was applied to integrate networks across descriptors. SNF iteratively diffused local similarity across network layers, producing a consensus representation that preserved shared structure while integrating complementary views. Its adaptation to molecular similarity networks followed the principles of data integration in complex systems.\cite{caldarelli2012networks} Robustness was evaluated under varying $k$NN values and repeated community detection.\cite{scalfani2022chemical}

\subsection{Statistical reporting and reproducibility}

All analyses were performed with fixed random seeds. Observed values were reported alongside null distributions with means, standard deviations, and empirical $p$-values. Core tools included RDKit,\cite{landrum2016rdkit} scikit-learn,\cite{pedregosa2011scikit} and NetworkX.\cite{hagberg2008exploring}

\bibliographystyle{apsrev4-2}
\bibliography{references}

@article{baldi2010computational,
  title={Computational approaches for drug design and discovery: An overview},
  author={Baldi, A},
  journal={Systematic reviews in Pharmacy},
  volume={1},
  number={1},
  pages={99},
  year={2010},
  publisher={Advanced Scientific Research}
}

@article{ashburn2004drug,
  title={Drug repositioning: identifying and developing new uses for existing drugs},
  author={Ashburn, Ted T and Thor, Karl B},
  journal={Nature reviews Drug discovery},
  volume={3},
  number={8},
  pages={673--683},
  year={2004},
  publisher={Nature Publishing Group UK London}
}

@article{lamb2006connectivity,
  title={The Connectivity Map: using gene-expression signatures to connect small molecules, genes, and disease},
  author={Lamb, Justin and Crawford, Emily D and Peck, David and Modell, Joshua W and Blat, Irene C and Wrobel, Matthew J and Lerner, Jim and Brunet, Jean-Philippe and Subramanian, Aravind and Ross, Kenneth N and others},
  journal={science},
  volume={313},
  number={5795},
  pages={1929--1935},
  year={2006},
  publisher={American Association for the Advancement of Science}
}

@article{barabasi2009scale,
  title = {Scale-free networks: A decade and beyond},
  author = {Barab{\'a}si, Albert-L{\'a}szl{\'o}},
  journal = {Science},
  volume = {325},
  number = {5939},
  pages = {412--413},
  year = {2009}
}

@article{barabasi2011network,
  title={Network medicine: a network-based approach to human disease},
  author={Barab{\'a}si, Albert-L{\'a}szl{\'o} and Gulbahce, Natali and Loscalzo, Joseph},
  journal={Nature reviews genetics},
  volume={12},
  number={1},
  pages={56--68},
  year={2011},
  publisher={Nature Publishing Group UK London}
}

@book{barabasi2016network,
  title = {Network Science},
  author = {Barab{\'a}si, Albert-L{\'a}szl{\'o} and P{\'o}sfai, M{\'a}ria},
  year = {2016},
  publisher = {Cambridge University Press}
}

@book{battiston2018multiplex,
  title={Multiplex and multilevel networks},
  author={Battiston, Stefano and Caldarelli, Guido and Garas, Antonios},
  year={2018},
  publisher={Oxford University Press}
}

@article{blondel2008fast,
  title = {Fast unfolding of communities in large networks},
  author = {Blondel, Vincent D and Guillaume, Jean-Loup and Lambiotte, Renaud and Lefebvre, Etienne},
  journal = {Journal of Statistical Mechanics: Theory and Experiment},
  year = {2008},
  pages = {P10008}
}

@book{caldarelli2007scale,
  title={Scale-free networks: complex webs in nature and technology},
  author={Caldarelli, Guido},
  year={2007},
  publisher={Oxford University Press}
}

@book{caldarelli2012networks,
  title = {Networks: A Very Short Introduction},
  author = {Caldarelli, Guido and Catanzaro, Michele},
  year = {2012},
  publisher = {Oxford University Press}
}

@article{cimini2019statistical,
  title   = {The statistical physics of real-world networks},
  author  = {Cimini, Giulio and Squartini, Tiziano and Saracco, Fabio and Garlaschelli, Diego and Gabrielli, Andrea and Caldarelli, Guido},
  journal = {Nature Reviews Physics},
  volume  = {1},
  pages   = {58--71},
  year    = {2019},
 
}

@article{de2013mathematical,
  title={Mathematical formulation of multilayer networks},
  author={De Domenico, Manlio and Sol{\'e}-Ribalta, Albert and Cozzo, Emanuele and Kivel{\"a}, Mikko and Moreno, Yamir and Porter, Mason A and G{\'o}mez, Sergio and Arenas, Alex},
  journal={Physical Review X},
  volume={3},
  number={4},
  pages={041022},
  year={2013},
  publisher={APS}
}

@article{freeman1977set,
  title = {A set of measures of centrality based on betweenness},
  author = {Freeman, Linton C},
  journal = {Sociometry},
  volume = {40},
  pages = {35--41},
  year = {1977}
}

@article{graham1985history,
  title = {On the history of the minimum spanning tree problem},
  author = {Graham, Ronald L and Hell, Pavol},
  journal = {Annals of the History of Computing},
  volume = {7},
  number = {1},
  pages = {43--57},
  year = {1985}
}

@article{guney2016network,
  title={Network-based in silico drug efficacy screening},
  author={Guney, Emre and Menche, J{\"o}rg and Vidal, Marc and Bar{\'a}basi, Albert-L{\'a}szl{\'o}},
  journal={Nature communications},
  volume={7},
  number={1},
  pages={10331},
  year={2016},
  publisher={Nature Publishing Group UK London}
}

@inproceedings{hagberg2008exploring,
  title = {Exploring network structure, dynamics, and function using NetworkX},
  author = {Hagberg, Aric A and Schult, Daniel A and Swart, Pieter J},
  booktitle = {Proceedings of the 7th Python in Science Conference},
  pages = {11--15},
  year = {2008}
}

@article{hopkins2008network,
  title={Network pharmacology: the next paradigm in drug discovery},
  author={Hopkins, Andrew L},
  journal={Nature chemical biology},
  volume={4},
  number={11},
  pages={682--690},
  year={2008},
  publisher={Nature Publishing Group US New York}
}

@article{jaccard1912distribution,
  title={The distribution of the flora in the alpine zone. 1},
  author={Jaccard, Paul},
  journal={New phytologist},
  volume={11},
  number={2},
  pages={37--50},
  year={1912},
  publisher={Wiley Online Library}
}

@article{kruskal1956shortest,
  title = {On the shortest spanning subtree of a graph and the traveling salesman problem},
  author = {Kruskal, Joseph B},
  journal = {Proceedings of the American Mathematical Society},
  volume = {7},
  number = {1},
  pages = {48--50},
  year = {1956}
}

@article{lancichinetti2012consensus,
  title = {Consensus clustering in complex networks},
  author = {Lancichinetti, Andrea and Fortunato, Santo},
  journal = {Scientific Reports},
  volume = {2},
  pages = {336},
  year = {2012}
}

@misc{landrum2016rdkit,
  title = {RDKit: Open-source cheminformatics},
  author = {Landrum, Greg},
  year = {2016},
  note = {https://www.rdkit.org}
}

@article{lipinski2012experimental,
  title={Experimental and computational approaches to estimate solubility and permeability in drug discovery and development settings},
  author={Lipinski, Christopher A and Lombardo, Franco and Dominy, Beryl W and Feeney, Paul J},
  journal={Advanced drug delivery reviews},
  volume={64},
  pages={4--17},
  year={2012},
  publisher={Elsevier}
}

@article{maslov2002specificity,
  title = {Specificity and stability in topology of protein networks},
  author = {Maslov, Sergei and Sneppen, Kim},
  journal = {Science},
  volume = {296},
  number = {5569},
  pages = {910--913},
  year = {2002}
}

@article{monti2003consensus,
  title = {Consensus clustering: A resampling-based method for class discovery and visualization},
  author = {Monti, Stefano and Tamayo, Pablo and Mesirov, Jill and Golub, Todd},
  journal = {Machine Learning},
  volume = {52},
  pages = {91--118},
  year = {2003}
}

@article{pedregosa2011scikit,
  title = {Scikit-learn: Machine learning in Python},
  author = {Pedregosa, Fabian and others},
  journal = {Journal of Machine Learning Research},
  volume = {12},
  pages = {2825--2830},
  year = {2011}
}

@article{rogers2010extended,
  title={Extended-connectivity fingerprints},
  author={Rogers, David and Hahn, Mathew},
  journal={Journal of chemical information and modeling},
  volume={50},
  number={5},
  pages={742--754},
  year={2010},
  publisher={ACS Publications}
}

@article{scalfani2022chemical,
  title = {Chemical space networks and visualization for molecular exploration},
  author = {Scalfani, Vincent F and others},
  journal = {Journal of Cheminformatics},
  volume = {14},
  pages = {1--15},
  year = {2022}
}

@article{sporns2013structure,
  title={Structure and function of complex brain networks},
  author={Sporns, Olaf},
  journal={Dialogues in clinical neuroscience},
  volume={15},
  number={3},
  pages={247--262},
  year={2013},
  publisher={Taylor \& Francis}
}

@article{szklarczyk2015string,
  title={STRING v10: protein--protein interaction networks, integrated over the tree of life},
  author={Szklarczyk, Damian and Franceschini, Andrea and Wyder, Stefan and Forslund, Kristoffer and Heller, Davide and Huerta-Cepas, Jaime and Simonovic, Milan and Roth, Alexander and Santos, Alberto and Tsafou, Kalliopi P and others},
  journal={Nucleic acids research},
  volume={43},
  number={D1},
  pages={D447--D452},
  year={2015},
  publisher={Oxford University Press}
}

@article{tabibzadeh2019myh9,
  title={MYH9-related disorders display heterogeneous kidney involvement and outcome},
  author={Tabibzadeh, Nahid and Fleury, Dominique and Labatut, Delphine and Bridoux, Frank and Lionet, Arnaud and Jourde-Chiche, Noemie and Vrtovsnik, Fran{\c{c}}ois and Schlegel, Nicole and Vanhille, Philippe},
  journal={Clinical Kidney Journal},
  volume={12},
  number={4},
  pages={494--502},
  year={2019},
  publisher={Oxford University Press}
}

@misc{tanimoto1958elementary,
  title={An elementary mathematical theory of classification and prediction},
  author={Tanimoto, Taffee T},
  journal={Technical Prepring},
  year={1958}
}

@article{veber2002molecular,
  title={Molecular properties that influence the oral bioavailability of drug candidates},
  author={Veber, Daniel F and Johnson, Stephen R and Cheng, Hung-Yuan and Smith, Brian R and Ward, Keith W and Kopple, Kenneth D},
  journal={Journal of medicinal chemistry},
  volume={45},
  number={12},
  pages={2615--2623},
  year={2002},
  publisher={ACS Publications}
}

@article{wang2014similarity,
  title = {Similarity network fusion for aggregating data types on a genomic scale},
  author = {Wang, Bo and others},
  journal = {Nature Methods},
  volume = {11},
  pages = {333--337},
  year = {2014}
}

@article{zhou2020network,
  title={Network-based drug repurposing for novel coronavirus 2019-nCoV/SARS-CoV-2},
  author={Zhou, Yadi and Hou, Yuan and Shen, Jiayu and Huang, Yin and Martin, William and Cheng, Feixiong},
  journal={Cell discovery},
  volume={6},
  number={1},
  pages={14},
  year={2020},
  publisher={Springer Singapore Singapore}
}
\end{document}